\documentclass[nofootinbib,twocolumn,preprintnumbers,superscriptaddress,aps,prl]{revtex4-1}
\usepackage{amsmath}
\usepackage{amssymb}
\usepackage{graphicx}
\usepackage{hyperref}
\usepackage{xspace,slashed}
\usepackage[utf8]{inputenc}

\usepackage{soul}

\usepackage{booktabs}
\AtBeginDocument{
\heavyrulewidth=.08em
\lightrulewidth=.05em
\cmidrulewidth=.03em
\belowrulesep=.65ex
\belowbottomsep=0pt
\aboverulesep=.4ex
\abovetopsep=0pt
\cmidrulesep=\doublerulesep
\cmidrulekern=.5em
\defaultaddspace=.5em
}

\makeatletter
\g@addto@macro\bfseries{\boldmath}
\makeatother

\newcommand*{\balancecolsandclearpage}{%
  \cleardoublepage
}

\newcommand{\tot}{\text{tot}}
\newcommand{\fm}{\;\mathrm{fm}}
\newcommand{\fb}{\;\mathrm{fb}}

\newcommand{\nb}{\;\mathrm{nb}}

\newcommand{\GeV}{\;\mathrm{GeV}}
\newcommand{\TeV}{\;\mathrm{TeV}}
\newcommand{\qhat}{{\hat q}}
\newcommand{\tp}{\text{top}}
\newcommand{\cQ}{{\cal Q}}
\newcommand{\lumi}{{\cal L}}
\newcommand{\reco}{\text{reco}}
\newcommand{\pttreco}{p_{t,\tp}^{\reco}}

\newcommand{\pbpb}{{\text{Pb}\text{Pb}}}

\usepackage{xcolor}
\definecolor{light-gray}{gray}{0.8}

\begin{document}

\preprint{CERN-TH-2017-237}

\newcommand{\CERNaff}{CERN, Theoretical Physics Department, CH-1211
  Geneva 23, Switzerland} 
\newcommand{\LisbonAffA}{LIP, Av. Prof. Gama Pinto, 2, P-1649-003 Lisboa , Portugal}
\newcommand{\LisbonAffB}{Instituto Superior T\'ecnico (IST), Universidade de Lisboa, Av. Rovisco Pais 1, 1049-001, Lisbon, Portugal}
\newcommand{\SantiagoAff}{Instituto Galego de F\'\i sica de Altas
  Enerx\'\i as (IGFAE),
  Universidade de Santiago de Compostela, Galicia-Spain}
\newcommand{\CNRSaff}{CNRS, UMR 7589, LPTHE, F-75005, Paris, France}

\title{Probing the time structure of the quark-gluon plasma with top quarks}
\author{Liliana Apolin\'ario}
\affiliation{\LisbonAffA}
\affiliation{\LisbonAffB}
\author{Jos\'e Guilherme Milhano}
\affiliation{\LisbonAffA}
\affiliation{\LisbonAffB}
\affiliation{\CERNaff}
\author{Gavin P.\ Salam}
\altaffiliation{On leave from \CNRSaff}
\affiliation{\CERNaff}
\author{Carlos A.\ Salgado}
\affiliation{\SantiagoAff}

\begin{abstract}
  The tiny droplets of Quark Gluon Plasma (QGP)
  created in high-energy nuclear collisions experience fast
  expansion and cooling  with a lifetime of a few $\text{fm}/c$. 
  Despite the information provided by probes such as jet quenching and
  quarkonium suppression, and the excellent description by
  hydrodynamical models, direct access to the time evolution of the
  system remains elusive.
  We point out that the study of hadronically-decaying $W$ bosons,
  notably in events with a top-antitop quark pair, can provide key novel
  insight, into the time structure of the QGP.
  This is because of a unique feature, namely a time delay between the
  moment of the collision and that when the $W$-boson decay products
  start interacting with the medium.
  Furthermore, the length of the time delay can be constrained by
  selecting specific reconstructed top-quark momenta.
  %
  %
  We carry out a Monte Carlo feasibility study and find that
  the LHC has the potential to bring first, limited information on the
  time structure of the QGP.
  Substantially increased LHC heavy-ion luminosities or future
  higher-energy colliders would open opportunities for more extensive
  studies.
\end{abstract}


\maketitle

The quark-gluon plasma (QGP), a state that characterised the first
microseconds of the universe, is regularly produced and studied in
ultrarelativistic heavy-ion collisions at both RHIC and the LHC.
A range of complementary probes is used to study the QGP.
These include properties that can be ascribed to hydrodynamic flow
patterns, suppression of heavy-quark bound states, hadrochemistry of
the final state, and modifications of the fragmentation of energetic
partons that traverse the medium (see e.g.\ \cite{Armesto:2015ioy}).
A property common to all these probes is that they are sensitive to
the properties of the QGP integrated over its lifetime.

Hydrodynamic simulation codes~\cite{Gale:2013da} predict a strong
time-dependence of the QGP's properties associated with its expansion
and cooldown, which last about $10\fm/c$ at the LHC.
It would be invaluable to develop a way of probing this
time-dependence.
The recent discovery (see e.g.\
\cite{Loizides:2016tew,Salgado:2016jws} and references therein) that
high-multiplicity proton--proton ($pp$) and proton--nucleus ($pA$) collisions
show signatures suggestive of
collective effects, in systems with significantly smaller lifetimes
than typical PbPb or AuAu collisions,
is an additional motivation for devising
a way of probing the time-structure of the QCD
medium. 

One powerful probe of the QGP is ``jet quenching'', i.e.\ the study of
modifications of jets that pass through the QGP (see
e.g. Ref.~\cite{Mehtar-Tani:2013pia}).
In all hard processes used so far for this purpose, dijet,
$\gamma$+jet or $Z$+jet production, the jets are produced
\emph{simultaneously} with the collision of the ions.

In this Letter, we point out that top-antitop ($t\bar t$) production
offers a unique novel opportunity to study the quark--gluon plasma, in
particular its time structure.
This is because, at variance with all other jet measurements
considered so far in the literature, the jets that come from the decay
products of the W-boson start interacting with the medium only at
\emph{later} times, due to a series of time delays.%
%
\footnote{In light
  of Ref.~\cite{Sirunyan:2017dnz}, similar measurements of the time
  structure of the QGP could be accessible with $W$+jet
  events. However, we will focus here on the $t\bar t$ avenue which we
  consider more promising (cf. the supplemental material).}
At rest, top quarks decay with a lifetime of about
$\tau_{top} \simeq 0.15\fm/c$ and the $W$ that is produced in the
top-quark decay has a lifetime of about $\tau_W \simeq 0.09\fm/c$.
When the $W$ boson decays hadronically, the resulting colour-singlet
quark-antiquark ($q\bar q$) pair is not immediately resolved by the
medium~\cite{CasalderreySolana:2012ef}.
Only after the $q$ and $\bar q$ have propagated and separated a
certain distance do they start interacting independently with the
medium.
We call this delay a decoherence time, $\tau_d$.
Thus the jets that are produced in the $t \to b + W \to q\bar q$ decay
chain do not see the full QGP, but only the part of the QGP that
remains after the sum of decay and decoherence times.
That sum of times is correlated to the momentum of the top quark, a
feature that may be exploited given a sufficient number of events.

To carry out a first investigation of the potential of using top
quarks for probing the time structure of the QGP, we proceed as
follows.
We take the average total delay time before the $W$ decay products start
interacting with the medium to be
\begin{equation}
  \label{eq:tau-total}
  \langle \tau_\tot \rangle = \gamma_{t,\tp} \tau_\tp 
        + \gamma_{t,W} \tau_W 
        + \tau_d\,,
\end{equation}
For the decay times, we use a transverse boost factor,
$\gamma_{t,X} = (p_{t,X}^2/m_X^2 + 1)^{\frac12}$, defined in terms
of the mass $m_X$, and transverse momentum $p_{t,X}$ of particle $X$.
The transverse component is the natural choice, because the frame in
which the top-quark has no longitudinal momentum is also the one in
which it is most natural to describe its interaction with the QGP,
which is approximately longitudinally-invariant.
We take the average  decoherence time to
be~\cite{CasalderreySolana:2012ef}
\begin{equation}
  \label{eq:taud}
  \tau_d = \left(\frac{12}{\hat q \theta_{q\bar q}^2} \right)^{1/3}\,,
\end{equation}
in natural units, $\hbar=c=1$, and with $\theta_{q\bar q}$
the opening angle between the two decay
products of the $W$, again in a longitudinal frame where the $z$
component of $W$ momentum is zero.
The quantity $\qhat$ is the transport coefficient of the medium
(squared transverse momentum broadening per unit length, see
e.g.~\cite{Mehtar-Tani:2013pia}).
While in practice it is expected to be a function of time, for our
proof of principle illustration here, we take it to be constant,
$\qhat = 4 \GeV^2 / \fm$~(conservatively taken larger than found in
Refs.~\cite{Burke:2013yra,Andres:2016iys}).
To get an event-by-event estimate of the interaction start time, we
will associate each component with a randomly distributed exponential
distribution.
With these choices, for inclusive top-quark production at the LHC
with centre-of-mass energy (per nucleon pair)
$\sqrt{s_{NN}} = 5.5\TeV$, the average times are, 
$\langle\gamma_{t,\tp} \tau_\tp\rangle \simeq 0.18 \fm/c\,$,
$\langle\gamma_{t,W} \tau_W\rangle \simeq 0.14 \fm/c\,$,
and $\langle\tau_d\rangle \simeq 0.34 \fm/c\,$,
with dispersions that are comparable.
The $1/3$ power in Eq.~(\ref{eq:taud}), means that
$\langle\tau_d\rangle$ is only weakly dependent on the value of
$\qhat$.

To probe jet quenching and its time dependence in $t\bar t$
production, we here suggest measuring the invariant mass $m_{jj}$ of
the dijet system that is 
produced from hadronic $W$ decays. 
In $pp$ events, $m_{jj}$ is closely related to the $W$ mass,
modulo final-state-radiation (FSR) effects.
The difference in reconstructed $m_{jj}$ in central ion-ion (AA)
collisions as compared to $pp$ will be our measure of jet quenching.

To evaluate the potential of such a study we examine
semi-muonic $t\bar t$ events, i.e.\ where one
top decays to $bW(W\to \mu\nu)$, while the other decays hadronically
to $bW(W\to jj)$.
In $pA$ collisions it has been demonstrated that it is possible to tag
this class of events with essentially no
background~\cite{Sirunyan:2017xku} as long as two $b$-tags are required,
and so we only consider signal events.

For a quantitative analysis, we use events from the ``\texttt{hvq}'' (heavy-quark)
process~\cite{Frixione:2007nw} in revision 3180 of the
\texttt{POWHEGbox}~\cite{Alioli:2010xd} generator, which
simulates top-quark production
to next-to-leading (NLO) accuracy in the
strong coupling constant.
We use it with the \texttt{PDF4LHC15\_nlo\_30} PDF
set~\cite{Butterworth:2015oua},
%
and shower events with
\texttt{Pythia}~8.223~\cite{Sjostrand:2006za,Sjostrand:2007gs},
tune 4C~\cite{Corke:2010yf}.
Our final results will be based on events at hadron level, without
underlying event.
The number of events that we can expect for an integrated luminosity
$\lumi_{AA}$ of $AA$ collisions is
$n(f) \simeq \lumi_{AA} \sigma_{pp}^{(t\bar t)} A^2 c(f)$
where $\sigma_{pp}^{(t\bar t)}$ is the $pp$ cross section for
$t\bar t$ production and $A$ is the atomic mass of the ions being
collided (see also Ref.~\cite{dEnterria:2015mgr}).
The $c(f)$ factor accounts for the centrality range $f$.
We will concentrate on $f = 0{-}10\%$, and so use
$c(0{-}10\%) \simeq 0.42$~\cite{Loizides:2017ack}.

To keep the analysis and simulation relatively simple, we choose not
to embed events in a heavy-ion medium.
Instead we introduce a single factor to mimic the combination of all
sources of fluctuations: those from the embedding and
medium-subtraction procedure, from finite detector resolution and also
from jet quenching dynamics.
Specifically, we rescale the momentum of each particle $i$ by a factor
$(1 + r \sigma_{p_t}/\sqrt{p_{t,i} + 1\GeV})$ where $r$ is a
Gaussian-distributed random number (different for each particle) with
a standard deviation of $1$; $\sigma_{p_t}$ is taken to be
$1.5\GeV^{1/2}$.
This leads to an effective relative jet energy resolution of about
$1.5\GeV^{1/2}/\sqrt{p_t}$ for high-$p_t$ jets, or about $15\%$ for
$p_t = 100\GeV$, consistent with Ref.~\cite{Sirunyan:2017jic}.

To simulate baseline full quenching in $0{-}10\%$ central PbPb systems, we
apply a constant energy loss rescaling factor $\cQ_0=0.85$ to all
particle momenta, which is consistent with observations in $\gamma/Z$+jets
measurements performed by ATLAS and CMS
\cite{ATLAS:2016tor,CMS:2016viv}.
Recall that the fluctuations associated with quenching are included in
our single global fluctuation factor.
A more sophisticated analysis would be possible, but is perhaps best
carried out in the context of a full experimental study.

To account for dependence of the quenching on the time $\tau_\tot$ at
which the $W$ decay products start to interact with the medium, all
particles from the $W$ decay are scaled by a factor $\cQ(\tau_\tot)$ rather
than $\cQ_0$.
We will return to the exact form of $\cQ(\tau_\tot)$ below.

To tag potential $t\bar t$ events, we require the presence of a muon,
two b-tagged jets and at least two non-b-tagged jets.
The muon should have $p_t > 25\GeV$ and rapidity $|y| < 2.5$.
Jets are obtained using the anti-$k_t$ jet
algorithm~\cite{Cacciari:2008gp} with radius $R=0.3$ and subsequent
partial declustering~\cite{Seymour:1993mx,Butterworth:2002tt} with the $k_t$
algorithm~\cite{Catani:1993hr,Ellis:1993tq}, all performed within
FastJet v3.2.1~\cite{Cacciari:2011ma}.
A selection requirement of $p_t > 30\GeV$ and $|y|<2.5$ is applied to the
anti-$k_t$ jets.
We assume a $b$-tagging efficiency of $\epsilon_b = 70\%$ per $b$, as
obtained in pPb events in \cite{Sirunyan:2017xku} and anticipating the
expected improvements in $b$-tagging in high-multiplicity environments
from HL-LHC detector
upgrades~\cite{CERN-LHCC-2015-020,Contardo:2020886}.
We also assume that fake $b$-tags do not introduce any substantial
background.

Our $W$ and top-quark reconstruction procedure is inspired by the
pseudo-top definition of Refs.~\cite{Aad:2015eia,CMSnote:pseudotop},
adapted to be more resilient to the presence of additional jets from
initial-state radiation (ISR) and at the same time robust with respect
to effects of quenching on the energy scales of $W$ and top
candidates.
The use of $R=0.3$ anti-$k_t$ clustering and then $k_t$ declustering to
obtain the input jets helps ensure adequate performance across a broad
range of top-quark transverse boosts (similar in spirit though
different in its details to Ref.~\cite{Gouzevitch:2013qca}).
The full procedure is detailed in the supplemental material.

\begin{figure}
  \centering
  \includegraphics[width=0.8\linewidth]{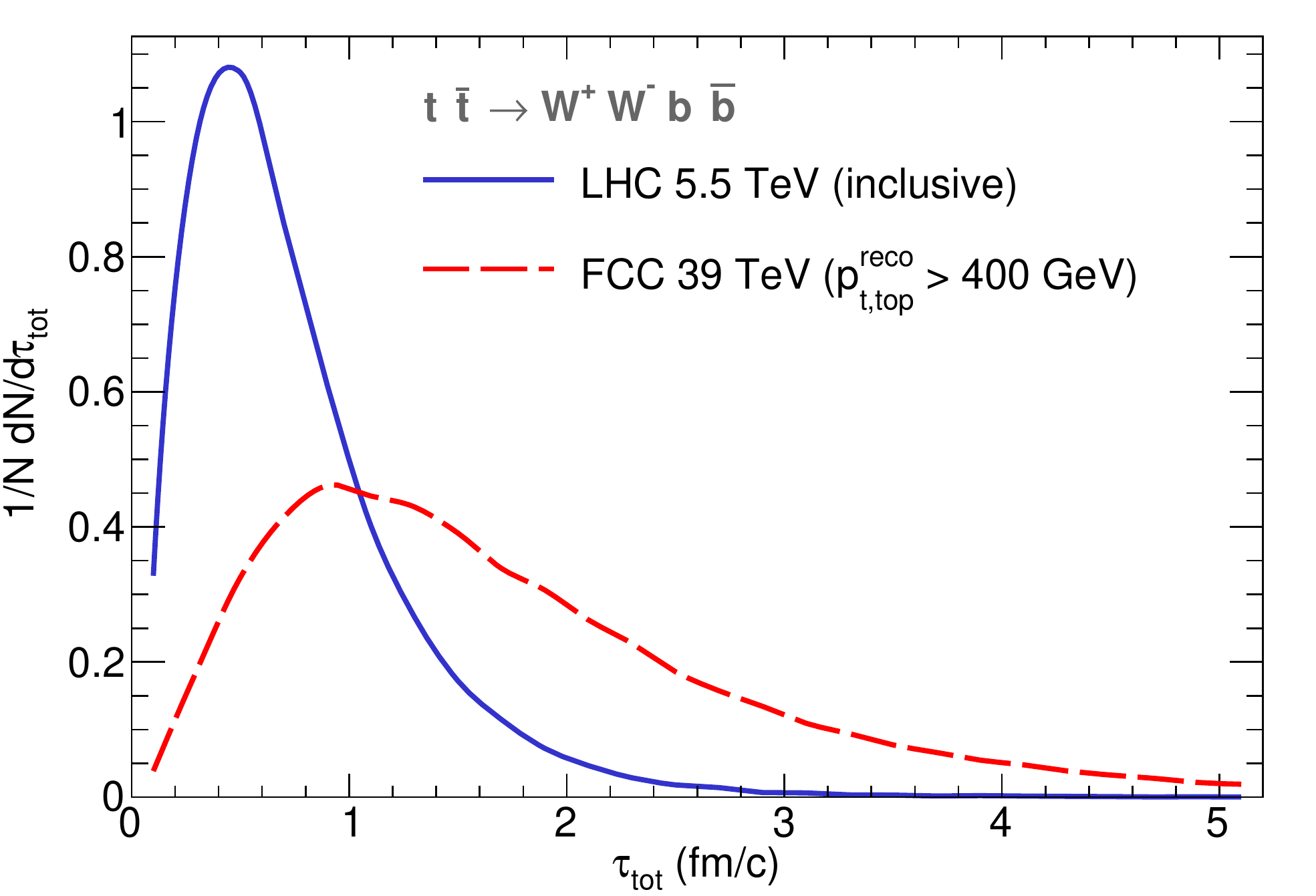}
  \caption{Distribution of $\tau_\tot$ for events that pass all
    reconstruction cuts and have a top-quark candidate (independently
    of the reconstructed top-quark and $W$-boson masses).
  } 
  \label{fig:times-dist}
\end{figure}

For each event that satisfies the reconstruction requirements, we
consider two observables: $m_{W}^\reco$, the mass of the reconstructed
hadronic $W$-boson candidate and $p_{t,\tp}^\reco$, the $p_t$ of the
corresponding top candidate.
The former will provide our measure of quenching (and was once before
studied for this purpose~\cite{Bhattacharya:2012gy}).
The latter can be translated
to an average $\tau_\tot$ and 
for $200\GeV\lesssim p_{t,\tp}^\reco \lesssim 1\TeV$ the relation reads
(see figure \ref{fig:times-v-pt} in the supplemental material)
\begin{equation}
  \label{eq:3}
  \langle \tau_\tot \rangle(p_{t,\tp}^\reco)
  \simeq
  (0.37 + 0.0022\, p_{t,\tp}^\reco / \text{GeV})\fm/c\,.
\end{equation}
The distribution of $\tau_\tot$ values is given in
Fig.~\ref{fig:times-dist} for the LHC $\sqrt{s_{NN}}=5.5\TeV$,
inclusively over $p_{t,\tp}^\reco$, and for a future-circular-collider
(FCC) with $\sqrt{s_{NN}}=39\TeV$, considering events with
$p_{t,\tp}^\reco > 400\GeV$.
Note the long tails in both cases, which will contribute sensitivity
to times substantially beyond $\langle \tau_\tot \rangle$.

\begin{figure}
  \centering
  \includegraphics[width=0.45\textwidth]{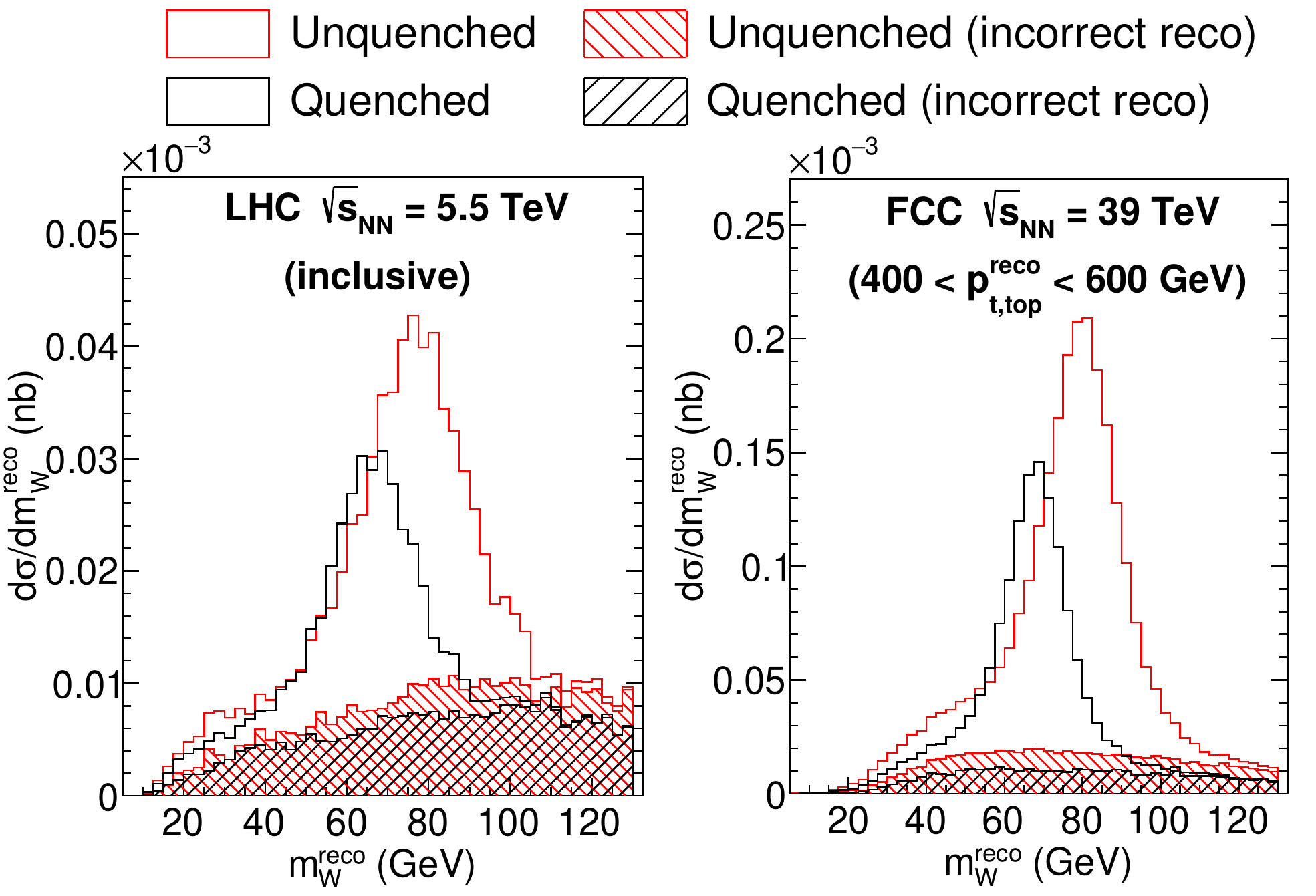}
  \caption{Differential fiducial proton--proton $t\bar t$
    reconstruction cross section as a function of $m_{W}^\reco$ at the
    LHC and FCC.  }
  \label{fig:reco-W-top}
\end{figure}

Fig.~\ref{fig:reco-W-top} shows the distribution of $m_{W}^\reco$,
again for the LHC and FCC, with a $p_{t,\tp}^\reco$ cut in the latter
case.
Results are shown with baseline full quenching for all particles and
without quenching (the latter being equivalent to $pp$ events embedded
in heavy-ion events to account for the effect of the underlying
event).
One sees clear $W$-mass peaks, superposed on a continuum associated
with events where the $W$ decay jets have not been correctly
identified.
The continuum is significantly reduced at high $p_{t,\tp}^\reco$.
The $W$ peaks in the quenched case are shifted to the left, and the extent
of the shift provides an experimental measure of the quenching.
The peaks are also lower in the quenched case, reflecting the smaller
fractions of events that pass the reconstruction (and, for FCC,
$p_{t,\tp}^\reco$) cuts.

To estimate the sensitivity of top-quark measurements to the
time-dependence of quenching in the medium, we consider a toy model in
which the quenching is proportional to the time between the moment
when the $W$ decay products decohere, $\tau_\tot$, and a moment when the
medium quenching effect stops being active, $\tau_m$.
This gives a $\tau_\tot$-dependent quenching factor $\cQ(\tau_\tot)$ for the $W$
decay products of
\begin{equation}
  \label{eq:time-dep-quenching-simple}
  \cQ(\tau_\tot) = 1 + (\cQ_0 - 1) \frac{\tau_m - \tau_\tot}{\tau_m} \Theta(\tau_m - \tau_\tot)\,.
\end{equation}
Recall that all other hadronic particles undergo quenching with the
factor $\cQ_0$.

For each choice of $\tau_m$ we obtain a $m_W^\text{reco}$ histogram
as in Fig.~\ref{fig:reco-W-top}.
We carry out a binned likelihood fit for the histogram and the
background of incorrectly reconstructed $W$'s using the functional
form
\begin{equation}
  \label{eq:fit}
  N(m) = a \exp \left[-\frac{(m-m_{W}^\text{fit})^2}{2\sigma^2}\right]
  + b + c\, m\,,
\end{equation}
which yields good fits.
The free parameters $a$, $b$, $c$, $\sigma$ and $m_W^\text{fit}$ are
constrained to sensible ranges so as to increase the stability of the
fit in low statistics samples.

\begin{figure}
  \centering
  \includegraphics[width=\columnwidth]{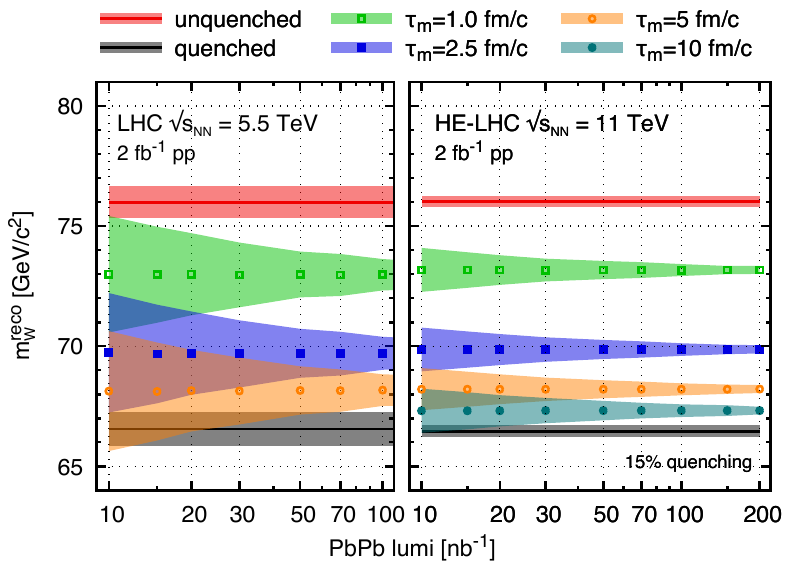}
  \caption{The average (points) and standard deviation (width of band)
    for $m_W^\text{reco}$ across many pseudo-experiments, as a
    function of luminosity for an inclusive sample of $t\bar t$
    events, as a function of the integrated PbPb luminosity at the LHC
    (left) and the HE-LHC (right).}
  \label{fig:Wreco-allPt-v-lumi}
\end{figure}

Fig.~\ref{fig:Wreco-allPt-v-lumi} shows the results for
$m_W^\text{fit}$.
They are plotted as bands for different $\tau_m$ values, as a function
of the PbPb integrated luminosity, $\lumi_\pbpb$.
The width of each band represents the standard deviation of
$m_W^\text{fit}$ values that we obtain when we carry out fits for a
large number of replica pseudo-experiments.
Two of the bands are independent of the PbPb luminosity: the top,
unquenched band, corresponds to the result that would be obtained by
embedding $2\fb^{-1}$ of $pp$ (unquenched) data into minimum-bias PbPb
events.
The bottom band is obtained by a similar procedure, but with the $pp$
jets' particles simply scaled down by the quenching factor $\cQ_0$,
i.e.\ by the quenching factor that would be expected if the $W$ decay
products were present and started interacting from time $0$.
In a real experiment, the corresponding scaling factor could be
obtained by measuring quenching in another quark-jet dominated process
(e.g.\ with $\gamma$+jet or $Z$+jet balance), as a function of the jet
$p_t$.

For short values of the effective medium lifetime, $\tau_m$, the
$m_W^\text{fit}$ result is close to the unquenched result.
This reflects the fact that the $W$ decay products start interacting
only towards the end of the medium lifetime.
For larger values of $\tau_m$ they instead still see most of the
medium duration, and most of the quenching.
A very short-lived medium, $\tau_m = 1\fm/c$, could be distinguished
from the full quenching baseline at the LHC with its currently
approved $\lumi_\pbpb = 10\nb^{-1}$.
However, to distinguish larger values of $\tau_m$ would require either
higher luminosities or higher energies.
This is illustrated in the right-hand plot of
Fig.~\ref{fig:Wreco-allPt-v-lumi} for a future HE--LHC
($\sqrt{s_{NN}} = 11\TeV$), where the $t\bar t$ cross section is $6$
times larger.

\begin{figure}
  \centering
  \includegraphics[width=\linewidth]{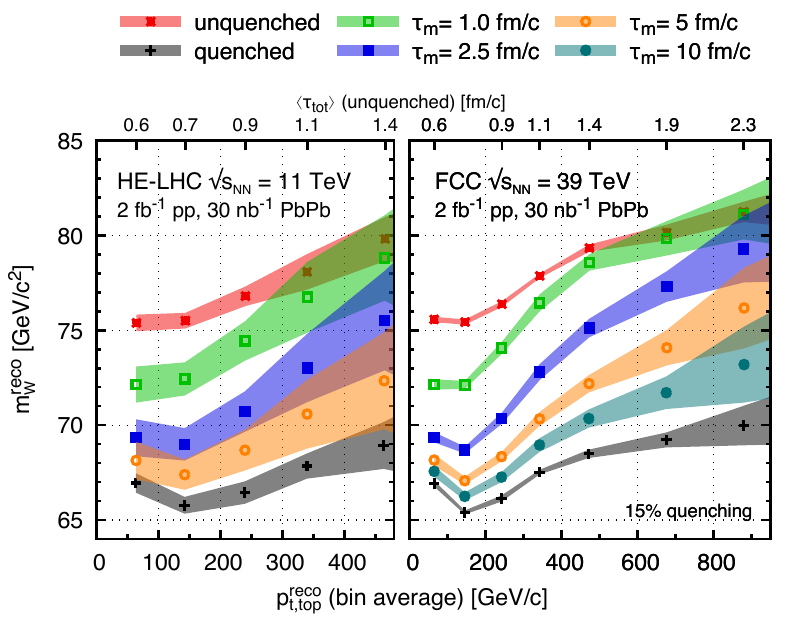}%
  \caption{Dependence of the reconstructed W mass on the 
    reconstructed top $p_t$ for HE-LHC (left) and FCC (right) collisions.
    The quenched result corresponds to baseline full modification of
    the $pp$ results, which would in practice be obtained using
    knowledge of quenching from other measurements.}
  \label{fig:Wmass-FCC39}
\end{figure}

At higher-energies it becomes advantageous to explore the
$\pttreco$ dependence of $m_W^\text{fit}$,
illustrated in Fig.~\ref{fig:Wmass-FCC39} for the HE--LHC
and the FCC ($\sqrt{s_{NN}} = 39\TeV$).
For each bin of $\pttreco$, the upper axis shows the corresponding
average $\tau_\tot$.
For a given band of $\tau_m$, when $\pttreco$ is large enough so that
$\langle \tau_\tot \rangle \gtrsim \tau_m$, the band merges with the
unquenched expectation.
Thus the shape of the $\pttreco$ dependence gives powerful information
on the medium time-structure.\footnote{The unquenched and
  baseline-quenched bands also have a $\pttreco$ dependence, induced
  by the underlying jet and muon $p_t$ cuts, as well as different
  amounts of final-state radiation outside the $R=0.3$ jet as a
  function of $\pttreco$.}

\begin{figure}
  \centering
  \includegraphics[width=\linewidth]{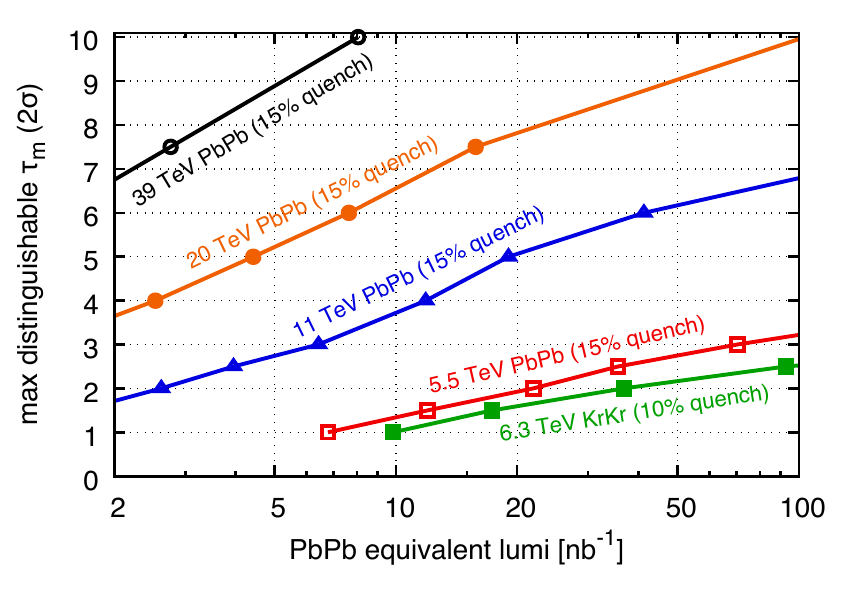}
  \caption{The maximum medium quenching end-time,
    $\tau_m$, that can be distinguished 
    from full quenching with two standard deviations, as a function of
    luminosity for different collider energies and species.
    For the KrKr points, the $\lumi_{KrKr}$ value that is used is
    equal to $\lumi_\text{PbPb} \cdot (A_\text{Pb}/A_\text{Kr})^2$,
    i.e.\ maintaining an equal number of nucleon--nucleon collisions.
    }
  \label{fig:2sigmatime-v-lumi}
\end{figure}

Fig.~\ref{fig:2sigmatime-v-lumi} shows our estimate of the maximum
$\tau_m$ that can be distinguished at two standard deviations from the
baseline full quenched result, for different colliders as a function of 
$\lumi_\pbpb$.
The number of standard deviations takes into account the statistical
uncertainty of $m_W^\text{fit}$, for both the actual heavy-ion data
and a reference sample as well as an
additional $1\%$ systematic uncertainty
(see supplemental material and
Refs.~\cite{CMS-PAS-FTR-17-002,ATLAS:2016tor}).
The reference sample is obtained using the same procedure as for the
bottom bands in Figs.~\ref{fig:Wreco-allPt-v-lumi} and
\ref{fig:Wmass-FCC39}, i.e.\ using $2\fb^{-1}$ of $pp$ events with a
rescaling of particle momenta by a factor $\cQ_0$ and inclusion of
underlying-event fluctuations.

%

%
For each collider luminosity and energy the results are obtained by
choosing a $\pttreco$ cut so as to maximise the significance.
We have verified that if we increase the fluctuations, $\sigma_{p_t}$,
the required luminosity scales as $\sigma_{p_t}^2$, in line with
expectations.



Lighter ions such as Kr are potentially promising, despite their smaller
quenching effects~\cite{Abelev:2009ab}, because of the potential for
order-of-magnitude higher effective integrated nucleon-nucleon
luminosities~\cite{Jowett:2017:HLHELHC,Bruce:2007mx}. They are
discussed further in the supplemental material.

To conclude, in this work we have shown that the study of top quarks
and their decays has a unique potential to resolve the time dimension
in jet-quenching studies of the QGP.
To benefit from this potential requires a sufficiently large sample of
top quarks, in particular to enhance event rates on the high-$p_t$
tail, which gives the sensitivity to the longer timescales.
At the LHC, with currently planned luminosity, such a programme could
begin.
With higher energy colliders or a significantly increased luminosity
at the LHC (whether from longer running or lighter ion species), there
would be substantial prospects for using jet quenching to study the
evolution of the QGP over the first few fm$/c$.
Overall, our results provide a strong motivation for a programme of
experimental studies of top-quark production in heavy-ion collisions.

\textbf{Acknowledgements:}
We are grateful to John Jowett for bringing to our attention the
potential for higher nucleon-nucleon luminosities with ions lighter
than Pb and to Detlef K\"uchler for information on possible ion
species in the LHC.
We would also like to thank Andrea Giammanco for exchanges regarding
pseudotop definitions, Phil Harris for discussions concerning the
use of $W$+jet events and Roberto Franceschini for a comment on the manuscript. 
This work was supported in part by the Funda\c{c}\~{a}o para a
Ci\^{e}ncia e Tecnologia (Portugal) under contracts
CERN/FIS-NUC/0049/2015 (LA and JGM),
Investigador FCT - Development Grant IF/00563/2012 (JGM)
and SFRH/BPD/103196/2014 (LA);
and by European Research Council grant HotLHC ERC-2011-StG-279579,
by Ministerio de Ciencia e Innovacion of Spain under project
FPA2014-58293-C2-1-P and Maria de Maetzu Unit of Excellence
MDM-2016-0692,
by Xunta de Galicia AGRUP2015/11 (CAS).

\bibliographystyle{apsrev4-1}
\bibliography{hi-top}


\balancecolsandclearpage

\onecolumngrid
\newpage
\onecolumngrid
\section{Supplemental material}
\twocolumngrid

\subsection{Using $W$+jet events}

One question that one might ask is whether one could use $W$+jet
events rather than $t\bar t$ events given that such events have two of
the sources of time delay: the $W$ decay and decoherence times.
It is especially natural to ask this question in light of the clear
hadronic $W$ peak observed recently for high-$p_t$ $W$+jet events by
the CMS collaboration~\cite{Sirunyan:2017dnz}.
However, we argue here that measuring $t\bar t $ looks more promising.

For $\sqrt{s} = 5.5\TeV$, the cross section for the $W$+jet process
with a $p_t$ cut of $\sim 100\GeV$ is comparable to the total
$t\bar t$ cross section.
In the case of semi-leptonic $t\bar t$ events, the 2 $b$ quarks,
lepton and (potentially) missing transverse momentum provide powerful
scope for tagging, albeit with a substantial price to pay in terms of
efficiency for the identification of all the decay products.

In the case of hadronic $W$+jet events there is a huge hadronic
background.
So far it has been possible to reduce this background sufficiently to
clearly see a $W$ (and $Z$) mass peak only at $p_t$'s of several
hundred GeV and with a $W$ tagging procedure that has moderate
efficiency, of the order of $10\%$.
This tagging procedure relies strongly on the pattern of radiation
from the $W$ decay products.
Any attempt to use similar methods for $W$ tagging in heavy-ion
collisions would have the drawback that they could bias the
reconstructed mass, for example by selecting $W$ events which do not
have additional medium-induced radiation.
The availability of tagging handles that are independent of the
hadronic $W$ in $t\bar t$ events thus makes the $t\bar t$ process
more attractive.

\subsection{Top reconstruction procedure}

Our reconstruction procedure requires as its starting point a muon,
two $b$-tagged jets and at least two non $b$-tagged jets.
Experimentally there would also be muon-isolation and possibly
missing-energy requirements, however these are not straightforward to
simulate correctly in heavy-ion collisions and we believe that
neglecting them here should not critically change the results.

One particularity of the reconstruction procedure is how we obtain the
input jets.
Firstly, we cluster all particles except the muon (and any neutrinos)
with the anti-$k_t$ jet finder~\cite{Cacciari:2008gp} (from FastJet
v3.2.1~\cite{Cacciari:2011ma}) with a radius of $R=0.3$.
Only jets with $p_t > 30\GeV$ and $|y|<2.5$ are accepted at this
stage.
For each anti-$k_t$ jet, its constituents are then reclustered with
the exclusive longitudinally-invariant $k_t$
algorithm~\cite{Catani:1993hr,Ellis:1993tq} with $R=1$ and
$d_\text{cut} = (20\GeV)^2$.
For low $p_t$ anti-$k_t$ jets this procedure usually yields just a jet
that is identical in particle content to the original anti-$k_t$ jet.
However for high $p_t$ anti-$k_t$ jets it can yield multiple exclusive
$k_t$ jets, notably associated with the substructure of boosted top
quarks and $W$ bosons.
This procedure is similar to the declustering approach of
Ref.~\cite{Seymour:1993mx} and to the $Y$-splitter algorithm of
Ref.~\cite{Butterworth:2002tt}, but applied to small-$R$ jets rather
than large-$R$ jets.
It is intended to provide the inputs for a top-reconstruction approach
that works at both low and high top-quark transverse boosts.
It is thus similar in its aims to the (technically different)
procedure of Ref.~\cite{Gouzevitch:2013qca}.

Given the muon, two $b$-tagged jets and at least two non $b$-tagged
jets obtained in this way, we use a procedure to identify the $W$ and
top-quark candidates that is in part inspired by the pseudo-top
definition of Refs.~\cite{Aad:2015eia,CMSnote:pseudotop}.
The $b$-jet that has the smallest distance to the muon is taken to
come from the leptonically decaying top ($b_\ell$, as opposed to
$b_h$).
The distance that is used is
$\Delta R_{b\mu}^2 = (y_b - y_\mu)^2 + (\phi_b - \phi_\mu)^2 $, where
are $y_i$ and $\phi_i$ are the rapidity and azimuthal angle of
particle $i$.

To reconstruct the hadronically decaying $W$ the standard pseudo-top
approach is to consider the two highest-$p_t$ non $b$-tagged jets.
However we found that this yielded poor reconstruction for high $p_t$
top quarks, because quite often one of the two highest-$p_t$ non
$b$-tagged jets comes from initial state radiation.
In full PbPb events this problem might be further exacerbated by jets
from other nucleon-nucleon interactions.

A widespread alternative approach (e.g.\
Ref.~\cite{CMSnote:pseudotop}) is to select and assign jets so as to
minimise a suitable kinematic fit variable.
This corresponds to choosing the subset of jets that is most
consistent with the known $W$ and top masses.
However, there is tendency for such a procedure to induce a peak 
around $m_W$ and $m_\tp$ in the mass distributions of incorrectly
reconstructed $W$ bosons and top quarks.
This complicates fits for the mass distribution of correctly
reconstructed (potentially quenched) $W$ candidates.

Ultimately the procedure that we adopted to identify the hadronic
$W$-decay jets was to consider all pairs ($j,k$) of non $b$-tagged
jets that satisfy $m_{jk} < 130\GeV$ and $m_{jkb_h} < 250\GeV$ and
select the pair with the largest $p_{t,j} + p_{t,k}$.
This pair is considered to be the reconstructed hadronically decaying
$W$ boson candidate, with mass $m_{W}^\reco \equiv m_{jk}$.
The reconstructed top-quark candidate is the combination of the $W$
candidate with the $b_h$ jet and its transverse momentum is denoted as
$p_{t,\tp}^\reco \equiv p_{t,jkb_j}$.
If no pair is found satisfying the above $m_{jk}$ and $m_{jkb_h}$ mass
conditions, the reconstruction is deemed to have failed and the event
is discarded.
From Fig.~\ref{fig:reco-W-top}, one sees that the mass distribution of
incorrectly reconstructed $W$'s is free of substantial shaping.

Our procedure, while adequate for the study of scenarios ranging from
the LHC to FCC, has not been the subject of extensive optimisation.
We believe that such an optimisation is better performed in the
context of a more detailed study that includes also full consideration
of all heavy-ion effects at a given specific collider.

\subsection{Contributions to the average total delay time, $\left\langle \tau_\tot \right\rangle$}

\begin{figure}[h]
  \centering
  \includegraphics[width=0.5\textwidth]{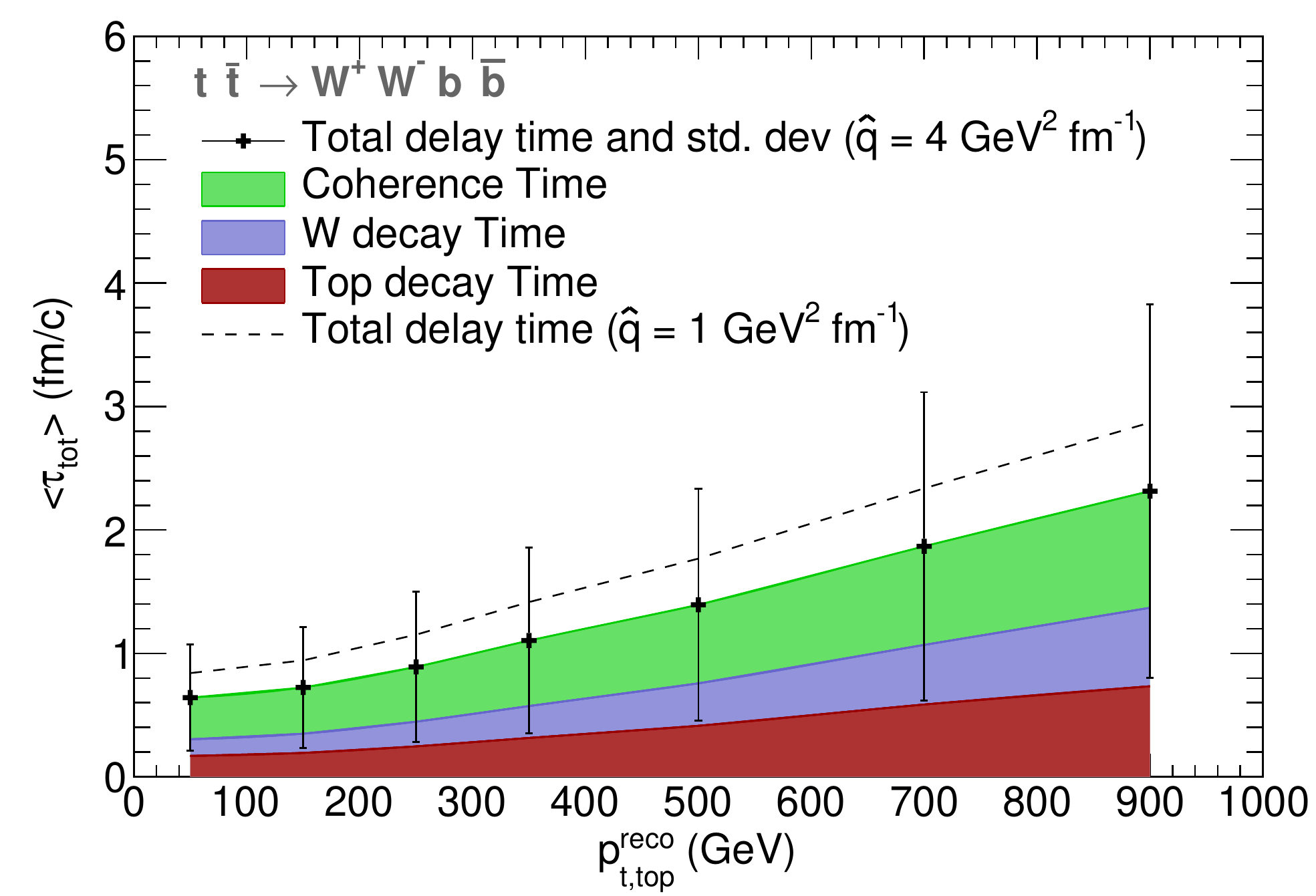}
  \caption{Total delay time and its standard deviation (markers
      and corresponding error bars), as given by
      Eq.~(\ref{eq:tau-total}), for $\hat{q} = 4$GeV$^2$/fm. The
      average contribution of each component is shown as coloured
      stacked bands (see legend). For comparison, the total delay time
      for $\hat{q} = 1$ GeV$^2$/fm is shown as a dashed line.} 
  \label{fig:times-v-pt}
\end{figure}

The result of Eq.~(\ref{eq:tau-total}) is shown as a function of the
reconstructed top jet transverse momentum in Fig.~\ref{fig:times-v-pt}, broken
into its three components, represented as stacked bands.
The range of $p_t$'s shown is guided by expectations as to what will be
accessible at widely discussed scenarios of potential future
colliders~\cite{Dainese:2016gch,Chang:2015hqa}.
The dispersion $\sigma_{\tau_\tot}$ of the sum of the three components is
also represented in Fig.~\ref{fig:times-v-pt}, as vertical black lines.
To illustrate the weak dependence of $\left\langle \tau_\tot \right\rangle$ on the 
value of $\hat{q}$, the average total delay time assuming
a $\hat{q} = 1 \GeV^2 / \fm$ (rather than $\hat{q} = 4  \GeV^2 / \fm$)
is shown as a dashed line. 
The larger result for $\tau_\tot$ would translate to a larger reach in
$\tau_m$ values for a given collider setup.

\subsection{Control of the jet energy scale}

To be able to identify the time-induced difference between quenching
of $W$ jets in $t\bar t$ events from full quenching, it is crucial to
have a reliable estimate of the expected reconstructed $W$ mass were
quenching of the $W$ jets to be unaffected by coherence delays and the
$W$ lifetime.

The procedure that we envisage for this purpose is to use measurements
of the $Z$-jet and $\gamma$-jet balance in events with cleanly
identified (leptonic) $Z$ bosons and photons to determine the
expectations for full quenching and to then apply that
determination to embedded $t\bar t$ events.

To estimate the potential precision of such an approach, we examined
how well the average $x_{jZ} = p_{tj}/p_{tZ}$ ratio could be determined at
the HL-LHC.
Ref.~\cite{CMS-PAS-FTR-17-002} from CMS gives a projection for the
uncertainties on the $x_{jZ}$ distribution with $\lumi_\text{PbPb} = 10\nb^{-1}$.
We took that distribution and created replica distributions by
fluctuating each bin with a Gaussian uncertainty set by the
projection.
We then evaluated the standard deviation of the $\langle
x_{jZ}\rangle$ values across many replicas.
The result for the standard deviation was $1.2\%$.
This guides our choice of $1\%$ for the systematic uncertainty on the
impact of standard quenching for the purpose of producing
Fig.~\ref{fig:2sigmatime-v-lumi}.

We also note that Ref.~\cite{ATLAS:2016tor} from ATLAS, shows a $1\%$
uncertainty (blue lines, bottom panel of Fig.3) for the
cross-calibration uncertainty between PbPb and $pp$ collisions.
One should keep in mind that other jet-energy scale uncertainties that
are common to the pp and PbPb cases should largely cancel when
considering the difference between embedded pp results and PbPb data
(and it is precisely this difference that interests us).

\subsection{Lighter ions}

Following the recent successful XeXe machine-development run at the LHC, the
prospect has been raised~\cite{Jowett:2017:HLHELHC} that with ions
lighter than Pb it might be possible to achieve effective
nucleon-nucleon luminosities (i.e.\ total number of hard collisions)
that are up to an order of magnitude larger than for PbPb, in part
because of the reduction of effects such as bound--free pair
production~\cite{Bruce:2007mx}. 
Generically, higher luminosities would bring substantially increased
sensitivity to the longer time structure of the QGP medium.

Aside from luminosity considerations, smaller ion species have both an
advantage and a disadvantage.
The advantage is that the intrinsic time scales associated with the
smaller, cooler QGP might be shorter than for PbPb and so more
accessible with top-quark probes.
However a smaller, cooler QGP is also likely to result in less
quenching.
It is for the purpose of illustrating the tradeoffs associated with
lighter species that in Fig.~\ref{fig:2sigmatime-v-lumi} we show a
curve labelled KrKr.
It uses a quenching of $10\%$ rather than $15\%$, in line with
observations in CuCu~\cite{Abelev:2009ab} that are consistent with quenching
that goes as $A^{1/3}$, where $A$ is the nuclear mass.
The reduced quenching means that the equivalent of
Fig.~\ref{fig:Wreco-allPt-v-lumi} for KrKr would have the bands more
closely spaced.
Accordingly one needs to go to higher luminosities in order to
distinguish any two given time scenarios.
At low luminosities the extra factor is relatively limited, about
$1.5$, while at higher luminosities it increases to about $3$.
Note that at higher luminosities the systematic and $pp$ statistical
uncertainties on the expected standard quenching results start to
dominate, since we have taken them to be independent of the PbPb
equivalent luminosity.

\end{document}